%% file: paper.tex
\documentclass{llncs}
\usepackage{cite}
\usepackage{amsmath}
\usepackage{amssymb}
\usepackage{graphicx}
\usepackage{subcaption}
\usepackage{listings}

\usepackage{braket}
\usepackage{tikz}
\usetikzlibrary{quantikz}

\newif\ifAnon
\Anonfalse  
\usepackage[linesnumbered,ruled,vlined,nosemicolon]{algorithm2e}
\usepackage{color}
\usepackage[pdfencoding=auto]{hyperref}

\renewcommand{\emph}[1]{\textit{\textbf{#1}}}
\let\epsilon\varepsilon

\begin{document}
\title{Quantum Tutte Embeddings}
\ifAnon
\author{Anonymous Author List}
\else
\author{
Shion Fukuzawa
\and
Michael T. Goodrich
\and
Sandy Irani
}
\institute{Dept. of Computer Science, Univ.~of California, Irvine, USA}
\fi

\maketitle
\begin{abstract}
Using the framework of Tutte embeddings, we begin an exploration
of \emph{quantum graph drawing}, which uses quantum computers to
visualize graphs.  The main contributions of this paper include
formulating a model for quantum graph drawing, describing how to
create a graph-drawing quantum circuit from a given graph, and
showing how a Tutte embedding can be calculated as a quantum state
in this circuit that can then be sampled to extract the embedding.
To evaluate the complexity of our quantum Tutte embedding circuits,
we compare them to theoretical bounds established in the classical
computing setting derived from a well-known classical algorithm for
solving the types of linear systems that arise from Tutte embeddings.
We also present empirical results obtained from experimental quantum
simulations.

\end{abstract}

\pagestyle{plain}

\input{intro}

\input{tutte}

\input{algs}

\input{experiments}

\ifAnon\else
\subsection*{Acknowledgements}
We would like to thank David Eppstein for helpful discussions regarding
the topics of this paper.
This work was supported in part by NSF grant 2212129.
\fi

\clearpage
    \bibliographystyle{splncs03}
    \bibliography{refs}

\clearpage
\begin{appendix}
\input{appendix}
\end{appendix}
\end{document}

%% file: intro.tex
\section{Introduction}
\emph{Quantum computing} studies ways to 
leverage the principles of quantum mechanics to perform computations;
see, e.g., Nielson and Chuang~\cite{nielsen2010}. Unlike
classical computing, where bits can only be in one of two states
(0 or 1), quantum computing uses quantum bits, or \emph{qubits}, which can
exist in multiple states at once, expressed as a  linear combination
of states with complex coefficients.
This property of qubits, called
\textit{superposition}, allows quantum computers to perform certain computations
faster than what is believed possible with classical computers.
Moreover, although
quantum computers are still in their infancy 
and face significant technological challenges, 
quantum computing holds tremendous promise for advancing fields 
ranging from chemistry and physics to artificial intelligence 
and machine learning.
At the very least, quantum computing is providing 
interesting alternatives to classical notions of
what is effectively computable.

As is well-known,
graphs are used to model a wide range of phenomena, including
molecular bonds, social network interactions,
biochemical pathways, and computer networks.
Fields that study such phenomena  include
promising applications of quantum computing; 
see, e.g.,~\cite{bassoli2021quantum,Cabello_2012,kumar2021quantum,matta2010quantum}.
In this paper, we therefore begin an exploration of
\emph{quantum graph drawing},
which studies how to use quantum computers to visualize graphs.
As a first step in this exploration, 
we focus in this paper on quantum circuits for what is 
arguably
the first graph 
drawing algorithm---\emph{Tutte embeddings}~\cite{tutte1963draw}.\footnote{
   Technically, proofs of F{\'a}ry's Theorem, showing 
   that simple, planar graphs have straight-line planar embeddings, came 
   earlier~\cite{wagner1936bemerkungen,istvan1948straight,stein1951convex},
   but these proofs, unlike Tutte's algorithm,
   don't provide vertex coordinates.}
As we review in this paper, a Tutte embedding is a type of
force-directed graph drawing paradigm that can be
viewed as a simulation of a physical system in which edges are 
represented as springs.
Intuitively, given a graph, $G$, the goal of this paradigm
is to find a layout that minimizes the 
total energy of the physical system defined by~$G$, after ``pinning''
some of the vertices in $G$ to specified locations, so as to
create an aesthetically pleasing visualization of~$G$; 
see, e.g.,~\cite{battista1999graph,kobourov2013}.

Although we focus on 
Tutte embeddings as a specific  
graph drawing paradigm, our goal in this paper is to formulate an approach
to quantum graph drawing that
could potentially be applied to other graph drawing paradigms as well.
That is, although
our approach uses a quantum linear systems solver to
compute the coordinates of the vertices in a graph embedding, we can
envision that other types of quantum circuits could be used for other
graph drawing paradigms, which could have similar data flows as the
methods we employ in this paper.

A quantum linear systems solver computes a vector $x$ that 
satisfies $A x  = b$ for input matrix, $A$, and vector, $b$. 
The matrix $A$ is embedded in the quantum circuit and the
vector $b$ must be prepared as a quantum state.
In a Tutte embedding, the matrix $A$ encodes spring-optimization equations
for the edges of a graph, so 
one of the first challenges we explore
is how to represent a graph in a quantum circuit. 
This requires developing a method for effectively 
encoding a graph in the gates of a quantum circuit, which
ideally satisfies some sparsity condition.
The output embedding is encoded in the vector $x$ as a quantum state,
so another issue that must be addressed is how to extract the drawing
from the quantum computer.

\paragraph{\textbf{Our Contributions.}}
Using the framework of Tutte embeddings, we provide in this paper a 
proof of concept for quantum graph drawing.
So as to avoid introducing additional complications in this endeavor,
we have not attempted to fully optimize the performance of the resulting 
quantum graph-drawing circuit.
Instead, the contributions
of this paper should be seen as providing a concrete example 
of an algorithm in quantum graph drawing, including formulating the model
and describing how to convert a graph into a quantum graph-drawing circuit,
which calculates a Tutte embedding as a quantum state that then
can be sampled to extract the embedding.
We compare the resulting complexity for our quantum Tutte embedding
circuit to theoretical bounds for computing such embeddings
in the classical computing setting
that can be derived from a well-known classical
algorithm for solving the types of linear systems that arise for
Tutte embeddings~\cite{spielman2004nearly}, 
as we review in the next section.
In addition, we also provide empirical results from
experimental quantum simulations.

%% file: tutte.tex
\section{Preliminaries}
In order for this paper to be as self-contained as possible,
we provide a brief primer on quantum computing in an appendix.

\paragraph{\textbf{A Brief Review of Tutte Embeddings.}}
In addition, let us provide
a brief review of Tutte embeddings~\cite{tutte1963draw}.
Let $G=(V,E)$ be an $n$-vertex simple, 3-connected planar graph. 
Let $f$ be a face of $G$, which we will consider to be the outer face.
A \emph{Tutte embedding} of $G$
is a crossing-free straight-line embedding of $G$ such 
that the outer face, $f$, is a convex polygon and such that each interior 
vertex (not on $f$)
is at the average (or barycenter) of its neighbors' positions. 
We enforce the first of these properties by ``pinning'' the vertices of $f$ to
be the vertices of a convex polygon, which is typically a regular convex
polygon.
We enforce the second property by imagining that the edges of $G$
are idealized springs with preferred length $0$, so that the minimum-energy
configuration of the interior vertices determines their positions.
In particular, we can formulate
two linear equations for each internal vertex, $u$,
of $G$
as follows (e.g., see the well-known graph-drawing book by
{Di~Battista}, Eades, Tamassia, and Tollis~\cite{battista1999graph}):
\[
\sum_{v \in N(u)} (x_u - x_v) = 0,
\]
and
\[
\sum_{v \in N(u)} (y_u - y_v) = 0,
\]
where $p_v=(x_v,y_v)$ is the point to which vertex $v$ is assigned, and
$N(u)$ denotes the set of neighbors of $u$ in $G$, i.e.,
$N(u)=\{v\,|\, (u,v)\in E\}$.
Note that for a vertex, $v$,
on the outer face, $f$, we pin $p_v=(x_v^*,y_v^*)$;
hence, $x_v^*$ and $y_v^*$ are fixed  constants.
As Tutte showed~\cite{tutte1963draw} (and which has been
repeated, e.g., 
by Hopcroft and Kahn~\cite{hopcroft1992paradigm} and
Floater~\cite{Floater}),
solving the above linear system of equations produces a
planar straight-line embedding of $G$ such
that every face is convex.

We subdivide $N(u)$ into the set, $N_I(u)$, of interior-neighbors of $u$,
and $N_f(u)$, neighbors of $u$ on the external-face, $f$.
That is,
\[
N_I(u)=\{v\,|\, (u,v)\in E \mbox{~and~} v\not\in f\}
\mbox{~~~and~~~}
N_f(u)=\{v\,|\, (u,v)\in E \mbox{~and~} v\in f\}.
\]
Also,
let $\deg(u)$ degree of the vertex, $u$.
Then we can rewrite the above linear equations for each interior
vertex, $u$, as follows:
\[
\deg(u) \,x_u - \sum_{v\in N_I(u)} x_v = \sum_{v\in N_f(u)} x_v^*,
\]
and
\[
\deg(u) \,y_u - \sum_{v\in N_I(u)} \,y_v = \sum_{v\in N_f(u)} y_v^*,
\]
where $p_v=(x_v^*,y_v^*)$ is the point to which the vertex $v$ is pinned if
$v\in f$.
Note that the righthand sides of the above equations are 
$0$ for each vertex, $u$, that is not adjacent to an outer-face vertex.
That is, this set of equations defines two linear systems,
\[
Ax =b_x,
\]
and
\[
Ay =b_y,
\]
where $x$ and $y$ are, respectively, the vector of $x$- and $y$-coordinates
of the points assigned to the interior vertices in $G$, 
and the entries of $b_x$ and $b_y$ are $0$ except
for vertices adjacent to pinned vertices.
Note further that the $(n-|f|)\times (n-|f|)$ matrix, $A$, is symmetric.
Moreover, it is closely related to 
the \emph{graph Laplacian}, $L$, for $G$, which is 
is defined as follows (see, e.g., \cite{chung1997spectral,pirani2015}):
\[
L = D - M,
\]
where $D$ is a diagonal matrix formed by the degrees of the vertices
in $G$ and $M$ is the adjacency matrix for $G$ (with the same ordering of 
vertices as in $D$).
Further, the matrix, $A$, is \emph{diagonally dominant}, that is,
$A_{i,i}\ge \sum_{j\not=i} |A_{j,i}|$, for all $i$.
Also, the number of non-zero entries in $A$ is $O(n)$, since the graph, $G$,
is planar.

Given such a symmetric, 
diagonally-dominate $O(n)\times O(n)$ matrix, $A$, with $O(n)$ non-zero entries,
and an $O(n)$-vector, $b$,
as well as an error tolerance, $\epsilon>0$,
Spielman and Teng~\cite{spielman2004nearly} 
provide a classical algorithm that
produces a vector, $\tilde x$, such that
$||A{\tilde x} - b||<\epsilon$ and $||{\tilde x}-x||\le \epsilon$,
where $x$ is the solution to $Ax=b$, in time
\[
O\left(n\log^{O(1)} n + n2^{O(\sqrt{\log n\log\log n})}
                         \log (\kappa/\epsilon)\right),
\]
where $\kappa$ is the \emph{condition number} of the matrix, $A$, i.e.,
the ratio of the largest to the smallest non-zero eigenvalue of $A$.
Thus, 
although the algorithm is fairly complicated and not commonly used in practice,
this provides a theoretical
classical algorithm for computing an $\epsilon$-approximation
to the coordinates of a Tutte embedding for $G$.
Incidentially, further improving the asymptotics for solving such
linear systems in the classical model is an on-going line
of research, with more recent results having similar ``near-linear''
bounds; see, e.g.,~\cite{kelner2013simple}.

Using
this result from the classical setting as a theoretical point of comparison,
our interest in this paper is to design a quantum circuit
for producing Tutte embeddings, based on a well-known
quantum linear system solver~\cite{harrow2009}, 
whose complexity can also be expressed as a function
of $n$, $\kappa$, and $\epsilon$, as well as $\Delta$, the maximum
degree of the vertices in $G$.
We have not attempted, however, to optimize the performance of the 
quantum linear system solver, as we mention above.
The contributions
of this paper should instead be seen as providing a first example 
for an algorithm in quantum graph drawing, including formulating the model
and describing how to convert a graph into a quantum graph-drawing circuit.
The quantum algorithm calculates a Tutte embedding as a quantum state which is then
sampled to extract the embedding.
We also provide some empirical results from
experimental quantum simulations.

%% file: algs.tex

\section{A Quantum Algorithm for Tutte Embeddings}



This section assumes basic familiarity with quantum computing. For
completeness, we provide a primer on quantum computing in an appendix.
Quantum algorithms for solving linear systems have been explored
heavily since being first introduced by Harrow et al.\cite{harrow2009},
who designed a quantum algorithm whose running time is exponentially
faster in the size of the matrix than classical methods, 
but with a few tradeoffs
that will be discussed later in this section. Their algorithm and
the follow-up work largely follow the same
structure, whose intuition can be captured by the following. Consider
a linear system, $A\ket{x} = \ket{b}$, and recall that the matrix
$A$ can be expressed in its eigenbasis as 
$A = \sum_{i} \lambda_i \ket{u_i} \bra{u_i}$. 
The inverse of $A$ can then be expressed as 
$A^{-1} = \sum_i \lambda_i^{-1} \ket{u_i} \bra{u_i}$, and the vector $\ket{b}$ can also
be written in this basis as $\ket{b} = \sum_i b_i \ket{u_i}$. Using
this, we can express $\ket{x}$ also in the eigenbasis of $A$
as follows:
\begin{align}
    \ket{x} &= A^{-1}\ket{b} \\
    &= \sum_i \frac{\beta_i}{\lambda_i} \ket{u_i}. \label{eq:ideal-x-state}
\end{align}
The objective of the quantum linear system solver is to prepare a state $\ket{x}$ which is proportional to the output state above. 
A vector corresponding to a quantum state must be normalized so that its $L_2$ norm is $1$, 
which is why the output is only proportional to  $\ket{x}$ given above,
which is not necessarily normalized.

Besides the dimensionality, $N$, of the matrix and the desired
precision of the solution, $\epsilon$, linear system solvers both
in the classical and quantum literature critically depend on two
other parameters, namely the sparsity, $s$, and the condition number,
$\kappa$. A matrix is \emph{$s$-sparse} if it has at most $s$ nonzero
entries in each row, and the condition number is defined to be the
ratio between the largest and smallest eigenvalue of $A$, i.e.,
$\kappa = \lambda_{max} / \lambda_{min}$. Intuitively, this captures
how ``invertible'' the matrix is, as a noninvertible matrix would
have $\kappa = \infty$. The following theorem captures the performance
of the quantum linear solver algorithm.  While there have been
subsequent developments in quantum linear systems 
solvers~\cite{childs2017}, they do not provide any advantage over the
original algorithm
for the linear systems we are considering.  
For a comprehensive survey on quantum
linear systems solvers, we refer the reader to 
Dervoric {\it et al.}~\cite{dervovic2018}.

\begin{theorem}[Harrow et al. \cite{harrow2009}]
\label{thm:harrow}
Let $A$ be a Hermitian $s$-sparse $N \times N$ matrix with condition
number $\kappa$ and let $b$ be a unit vector. Then there is a quantum
algorithm that can prepare the state, $\ket{x}$, satisfying the
equation
\begin{equation}
    Ax = b
\end{equation}
in $\tilde{O}\left(\kappa T_B + \log\left(N\right) s^2 \kappa^2/\epsilon \right)$ time. The $\tilde{O}$ suppresses the more slowly growing terms of $(\log^* (N))^2$, $\exp\left(O\left(1/\sqrt{\log(t_0, \epsilon_H})\right)\right)$, and polylog$(T/\epsilon_\Psi)$. Here, $T_B$ is the time required to prepare the input vector $\ket{b}$, $\epsilon_\Psi$ and $\epsilon_H$ are error terms accrued in the phase estimation subroutine, and $T$ is usually selected to be $O(\log (N) s^2 t_0)$. The $\epsilon$ is the additive error achieved in the output state $\ket{x}$. 

Once the output state $\ket{x}$ is prepared, for any Hermitian operator, $M$, the expected value $x^TMx$ can be measured.
\end{theorem}

As described above in section 2, the matrix $A$ that appears in the
Tutte embedding system is diagonally dominant, and for a fixed
degree graph has constant sparsity, $s$. It is easy to show that this
matrix is sparse when the degrees of the vertices are bounded by
$\Delta$, for example. The analysis of the condition
number is more involved, and we provide an experimental analysis
of how the condition number can scale for various classes of graphs.

Our quantum graph drawing algorithm has 3 main components: preparing the input state $\ket{b}$, the procedure for generating the solution vector $\ket{x}$, and finally the measurement of the expected value of an operator $M$ given the state $\ket{x}$. The following is an overview of the algorithm, and specific parts of it will be detailed in the following subsections. 

\begin{enumerate}
    \item Preparation
    \begin{itemize}
        \item Prepare the input vector $\ket{b}$.
        \item Generate a quantum circuit from a sparse representation of $A$. 
    \end{itemize}
    \item Run eigenvalue estimation 
    \item Perform conditional rotations 
    \item Invert the eigenvalue estimation 
    \item Repeat steps 2, 3, 4 until success
    \item Use output vector $\ket{x}$ to measure summary statistics about an operator $M$
\end{enumerate}

\subsection{Preparing the Vector $\ket{b}$}\label{subsect:preparing-b}

Recall that
for an $n$-vertex planar graph, we can draw the graph
by solving two system of equations, 
$F(u) = \sum_{(u,v) \in E} (v - u)$, 
for the $x$ and $y$ coordinates, respectively. 
Also recall that if $k$ vertices
are pinned down, the matrix $A$ representing the graph is
close to the graph Laplacian. Thus, the final linear systems
will be represented by a system of size $n - k$ by $n - k$. In the
literature, such matrices have been defined and studied as
``grounded Laplacians.'' For example,
Pirani et al. \cite{pirani2015} show that for weighted
$d$-regular graphs with one pinned vertex, the smallest
eigenvalue is $\Theta(1/n)$. We empirically study the condition
number for grounded Laplacians of several classes of graphs and
present the results below in section \ref{sec:experiments}.

\begin{figure}
    \centering
    \includegraphics[width=.8\linewidth]{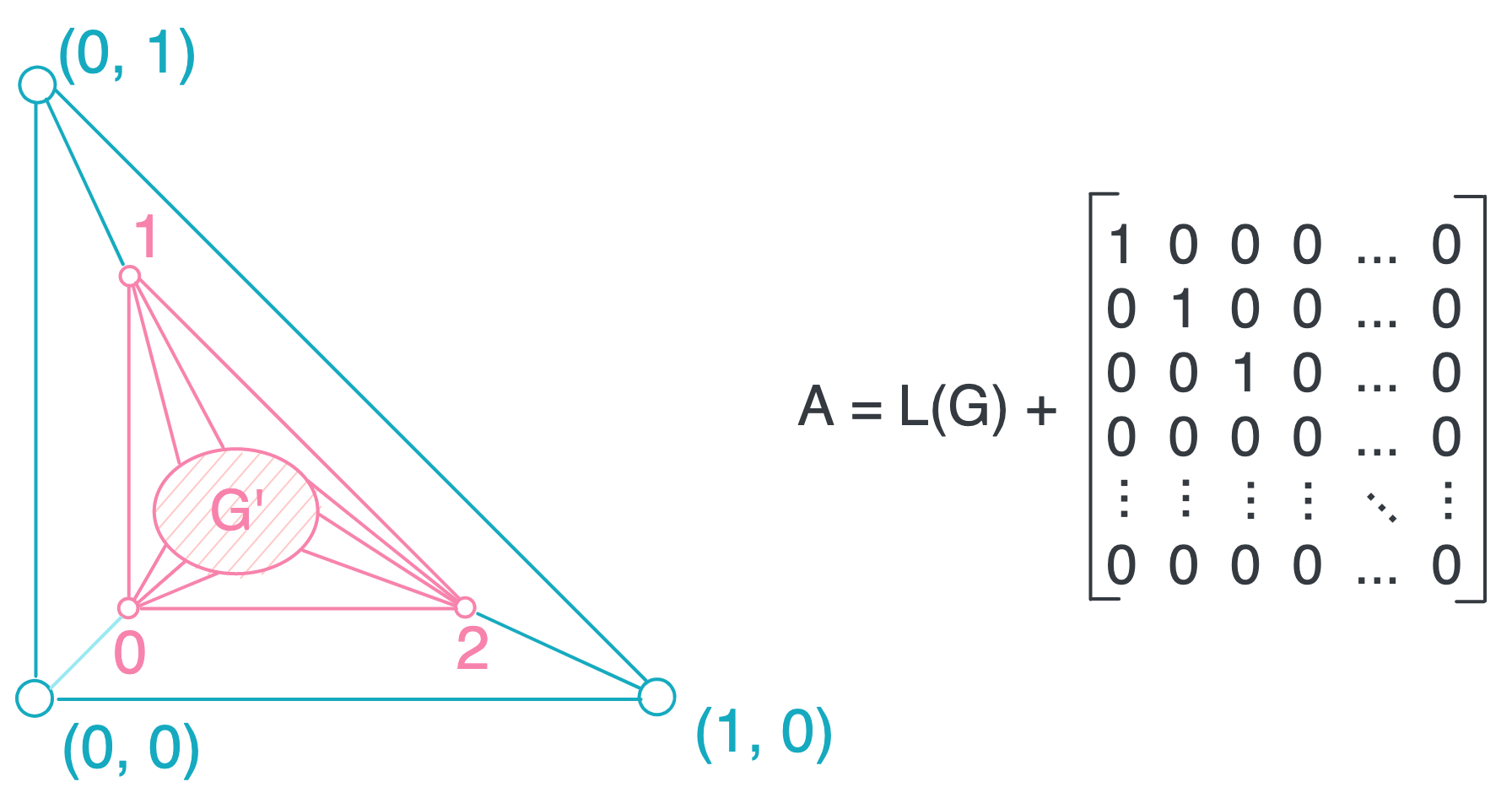}
    \caption{To simplify the linear system we are solving, we construct our system by creating a dummy outer face drawn in blue, and pinning the three blue vertices to the coordinates $(0, 0)$, $(0, 1)$, and $(1, 0)$. Under this construction, if we index the vertices of our graph $G$ such that the three outer vertices are indexed as above, the construction of the matrix $A$ is simply the graph Laplacian $G(L)$ plus a diagonal matrix such that the first three entries are 1, and the remaining are all 0. This indexing scheme also ensures that $\ket{b}$ has a single index with a 1 and 0 everwhere else. }
    \label{fig:dummy-construction}
\end{figure}

In this paper, we focus on the case where there is a triangular
outer face of the graphs, which will always be drawn such that it
bounds the rest of the graph. Furthermore, to simplify the input
vector, $\ket{b}$, we add a dummy outer face such that the true outer
face is connected to this dummy outer face, as is shown in 
Figure~\ref{fig:dummy-construction}. In our construction of the linear
system, we pin down this dummy outer face to the coordinates,
$(0, 0)$, $(1, 0)$, $(0, 1)$. For the purpose of illustration, when
indexing the vertices of the graph we will always index the vertices
of the outer face first in a clockwise order starting from the
bottom left vertex. When using this indexing scheme, the $\ket{b}$
vector for the $x$-coordinate system is always the vector with a 1
in index 2 and 0's everywhere else, and the $\ket{b}$ vector for
the $y$-coordinate system is always the vector with a 1 in index 1
and 0's everywhere else.

\subsection{Preparing the circuit to implement the matrix $A$}
Recall that quantum gates are unitary operators, 
and the input matrices are Hermitian. 
Given a Hermitian matrix, $A = \sum_i \lambda_i \ket{u_i}$, if we take the matrix exponent of $A$, then we get an operator $U$ which is unitary: 
\begin{equation}\label{eq:matrix-power}
    U := e^{-iAt} = \sum_i e^{-i \lambda_i t} \ket{u_i}\bra{u_i}.
\end{equation}
The factor of $i$ is critical in enforcing unitarity, and the constant multiple of $t$ is a parameter we vary during the algorithm. An important preparation step in running this algorithm on a quantum computer is to generate the circuit that implements $U$. 

In the literature, this problem is referred to as Hamiltonian simulation 
(See Dervovic {\it et al.}~\cite{dervovic2018} for a more in depth tutorial). 
The main method we use begins by decomposing our operator into a sum of simpler matrices as $A = \sum_{j=1}^s A_j$, where simple refers to a class of matrices that are known to be easy to implement on a quantum computer. Given this decomposition, equation \ref{eq:matrix-power} can be rewritten,
\begin{equation}\label{eq:matrix-sum-decomposition}
    e^{-iAt} = e^{-i\sum_{j=1}^s A_j t}.
\end{equation}
We want to combine smaller circuits where each circuit depends only on a single $A_j$, but in general matrix products do not commute so we cannot just implement the circuit $\prod_{j=1}^s e^{i A_j t}$. 
Instead, the Lie-Trotter product theorem~\cite{trotter1959product} states that 
\begin{equation}\label{eq:lie-product}
    e^{-i(A + B)t} = \lim_{m \rightarrow \infty} \left(e^{-i A t/m} e^{-iBt/m}\right)^m
\end{equation}
which has driven many results in Hamiltonian simulation. 
For the simplest model of approximating equation~(\ref{eq:lie-product}), it is now known that selecting $m = O((\nu t)^2 / \epsilon)$ achieves
\begin{equation}\label{eq:lie-product-approx-1}
    \left|\left| \left(e^{-i A t/m} e^{-iBt/m}\right)^m - e^{-i(A + B)t} \right|\right| \leq \epsilon,
\end{equation}
where $\nu := \max \{||A||, ||B||\}$ and $||\cdot||$ denotes the spectral norm of the operator.  

To decompose the operator $A$ into a sum of terms, we can first express it as the sum of the diagonal $D$ and non-diagonal $M$ which is just the adjacency matrix of the free vertices. 

A simple circuit construction for the diagonal matrix $e^{-iDt}$ can be realized by the following. This operation will use two quantum registers, the index register with $\log N$ qubits and a memory register using $\log s$ qubits, where $s$ is the sparsity. The circuit begins by setting the index register into superposition using $\log N$ Hadamard gates. Then for each index, it will perform a $\log N$ qubit Toffoli gate up to $\log s$ times to store the state $\ket{a, d(a)}$ where $d(a)$ is the diagonal element in the matrix of the $a$-th index. 
After this step, we apply phase gates parameterized by $-2^i t$ on the $i$-th qubit. 
After this step, we then uncompute the memory register by reversing the Toffoli gates. The full transformation of the above circuit is described by the following sequence of equations:
\begin{align}
    \ket{a, 0} &\rightarrow \ket{a, d(a)} \\
    &\rightarrow e^{-id(a) t} \ket{a, d(a)} \\
    &\rightarrow e^{-id(a) t} \ket{a, 0} \\
    &= e^{-iDt} \ket{a}\ket{0}
\end{align}

Now that we have a circuit computing the diagonal part, what remains
is to build a circuit to implement the exponentiated adjacency
matrix, $M$. Unfortunately, $M$ on its own may not be easily translated into a
quantum circuit, so this must also be expressed as a sum of terms
that can be easily diagonalized. For an adjacency matrix, this can
be accomplished by computing an edge coloring on the graph, and
decomposing the adjacency matrix into a sum of terms corresponding
to each color. Since the graphs we are considering have a maximum
degree of $\Delta$, a greedy edge coloring scheme provides a
coloring of at most $2\Delta$ terms. After this step, each color,
$c$, corresponds to a 1-sparse matrix that is still Hermitian,
meaning that it can easily be diagonalized to be written as 
$M_c = U_cD_cU_c^\dagger$ where $U_c$ is unitary and $D_c$ is diagonal.
Using the identity,
\begin{equation}
    e^{-iUHU^\dagger t} = Ue^{-iHt}U^\dagger,
\end{equation}
for conjugations of Hermitian matrices with unitary matrices, we can build the circuit $M_c$ by first constructing the circuit to simulate $U_c$, then performing the diagonal part the same way as it was done for $D$. 

In the following sections, we will use $U$ to refer to the circuit that 
implements the approximation to the $e^{-iAt}$ equation decomposed into a sum of a diagonal matrix and the decomposition of the off-diagonal matrix corresponding to the edge coloring, using the Trotterization scheme mentioned 
in equation~(\ref{eq:lie-product-approx-1}).

\subsection{Solving the Linear System}

\begin{figure}
    \centering
    \includegraphics[width=.9\linewidth]{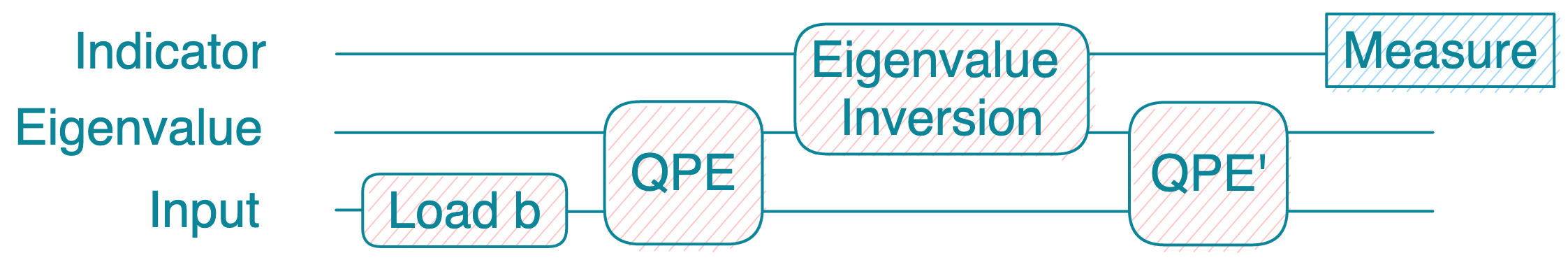}
    \caption{A high level overview of Harrow's quantum algorithm for solving linear systems. The shaded boxes describe subroutines that will be described in further detail in this section.}
    \label{fig:hhl-overview}
\end{figure}

In this section, we provide a high-level overview 
of how Harrow et al.'s linear system solver algorithm~\cite{harrow2009} works. 
The first key subroutine is what is referred to in the literature as phase estimation or eigenvalue estimation \cite{kitaev95, nielsen2010}. Since phase estimation is a standard quantum algorithm, we just describe the input/output behavior of the algorithm and how it is used for linear system solving.


The ideal phase estimation algorithm takes as input a unitary $U$ and one of its eigenvectors $\ket \psi$, and the output will be an $r$-bit approximation of $2^{r}\theta$ for $e^{2\pi i \theta}$, the eigenvalue corresponding to the input eigenvector. The procedure is described by the equation,
\begin{equation}
    \ket{0}_r \ket{\psi} \rightarrow \ket{2^{r}\theta} \ket{\psi}.
\end{equation}


The behavior of the algorithm on a generic quantum state $\ket{b}$ can be understood
by expressing $\ket{b}$ in the eigenbasis for $U$.
Let $\ket{\psi_j}$ denote the $j$-th eigenvector of $U$. Then $\ket{b}$  can be expressed as  $\ket{b} = \sum_j \beta_j \ket{\psi_j}$ for some set of coefficient $\{\beta_j\}_j$.
Performing the full phase estimation using this state as the input state now prepares a superposition of $r$-bit approximations of $2^r \theta_j$ where $e^{2\pi i \theta_j}$ is the eigenvalue of $\ket{\psi_j}$ weighted by the coefficient $\beta_j$:

\begin{equation}
    \ket{0}_r \ket{b} = \sum_j \beta_j \ket{0}_r \ket{\psi_j} \rightarrow \sum_j \beta_j \ket{2^{r}\theta_j} \ket{\psi_j}
\end{equation}

Note that $2 \pi \theta_j = \lambda_j$ for all $j$, and a uniform scaling of the eigenvalues will not effect the description of the algorithm below, which is expressed in terms of the $\lambda_j$'s.
Once we've prepared the register storing a superposition of the eigenvalues, we use these to prepare the following state:
\begin{equation}\label{eq:controlled-rotation}
    \sum_j \beta_j \ket{\lambda_j}\ket{\psi_j} \left( \sqrt{1 -   \frac{C^2}{\lambda_j^2}}\ket{0} + \sqrt{\frac{C}{\lambda_j}}\ket{1}\right),
\end{equation}
where $C$ is a constant that should be chosen such that $|C| < \lambda_{min}$. This rotation can be performed by preparing a single qubit register initialized in the $\ket{0}$ state, then for each $k$-th significant bit in the eigenvalue register, perform a controlled rotation by an angle of $\frac{C}{2^k}$. Provided a sufficiently good approximation of the eigenvalue, this will put the quantum device in the state described in equation~(\ref{eq:controlled-rotation}). 

Following the controlled rotation, we now do the inverse of the phase estimation procedure, effectively \textit{uncomputing} the eigenvalue register. We have no entanglement with the eigenvalue register now, so we can ignore it for the remaining analysis. The state we have now is 
\begin{equation}\label{eq:uncomputed-state}
    \sum_j \beta_j \ket{\psi_j} \left( \sqrt{1 -   \frac{C^2}{\lambda_j^2}}\ket{0} + \sqrt{\frac{C}{\lambda_j}}\ket{1}\right),
\end{equation}
and the final step before preparing the $\ket{x}$ vector is to measure the last register. If we measure $\ket{1}$ in the final register, we end up with the state 
\begin{equation}\label{eq:post-measurement-state}
    \sqrt{\frac{1}{\sum_j C^2 |\beta_j|^2 / |\lambda_j|^2}} \sum_j \frac{\beta_j C}{\lambda_j} \ket{\psi_j},
\end{equation}
which is exactly the state $\ket{x}$ described in 
equation~(\ref{eq:ideal-x-state}) up to a normalization factor. The final running time including the time required to perform the matrix to circuit conversion is summarized in the following theorem. 

\begin{theorem}
Let $G$ be a 3-connected planar graph with $n$ vertices and at most $\Delta$ edges per vertex. Let $A$ be the grounded Laplacian and $\ket{b_x}$, $\ket{b_y}$ be the input vectors for the $x$- and $y$- coordinate systems as described in section 3.1. Then, the quantum computer can prepare the normalized solution vector $\ket{x}$ in time $\tilde{O}\left( sn + \kappa T_B + \kappa^2s^2 \log (n) / \epsilon \right)$, where the first term is the time required to construct the quantum circuit from the input matrix. 
\end{theorem}

%% file: experiments.tex
\section{Experiments}\label{sec:experiments}

This section is divided into two sections, the first being an empirical study on the condition number of grounded Laplacian matrices, and the second section being a demonstration of the results obtained from using the quantum algorithm for graph drawing. 

\subsection{Condition Number}

\begin{figure}
    \centering
    \includegraphics[width=0.8\textwidth]{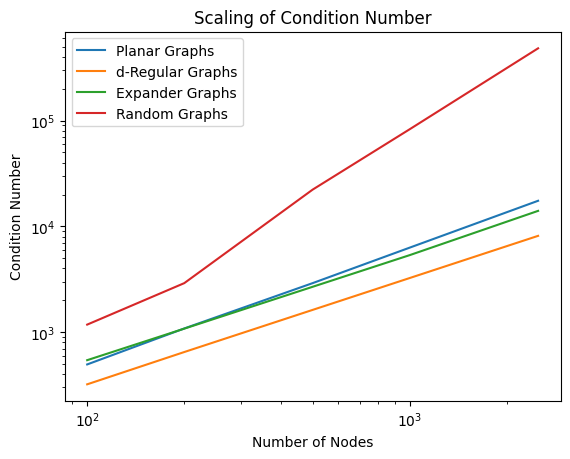}
    \caption{An empirical study of the scaling of condition numbers for standard classes of graphs. Graphs were generated randomly within each class, and the average is plotted over 100 runs per data point. }
    \label{fig:condition-number-scaling}
\end{figure}

As the quantum algorithm critically relies on the condition number, we begin the exploration with an empirical study of these values for a few classes of graphs. The planar graphs were generated using the technique outlined in the following subsection. The remaining three types of graphs were generated using the built-in graph generation functions for Python's NetworkX library. The expander graphs are Margulis-Gabber-Galil graphs and the random graphs 
are Erd\H{o}s-R{\'e}nyi graphs. For each instance, the average over 100 samples is plotted in Figure \ref{fig:condition-number-scaling}. We see in the results that for all chosen classes of graphs, the condition number appears to scale at least linearly. 

\subsection{Example Quantum Tutte Embeddings}

\begin{figure}
     \centering
     \begin{subfigure}[b]{0.49\textwidth}
         \centering
         \includegraphics[width=\textwidth]{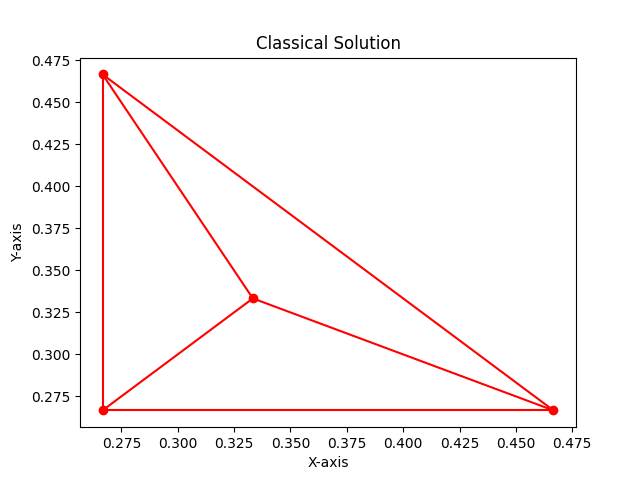}
     \end{subfigure}
     \hfill
     \begin{subfigure}[b]{0.49\textwidth}
         \centering
         \includegraphics[width=\textwidth]{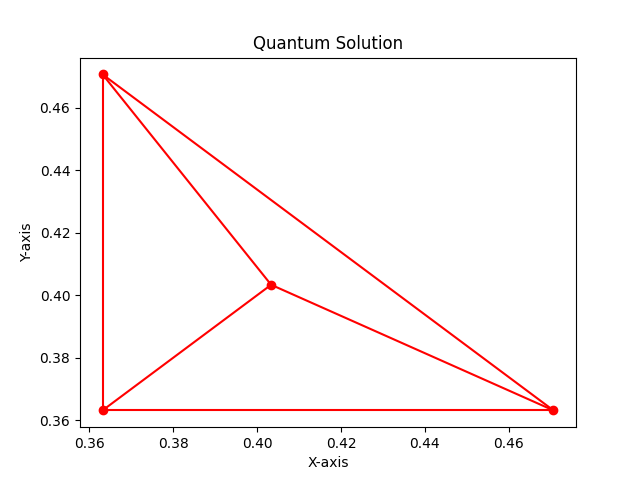}
     \end{subfigure}
     \hfill
     
     \begin{subfigure}[b]{0.49\textwidth}
         \centering
         \includegraphics[width=\textwidth]{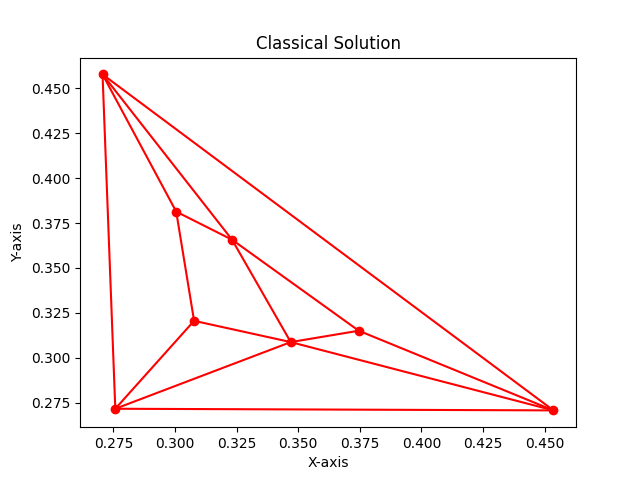}
     \end{subfigure}
     \hfill
     \begin{subfigure}[b]{0.49\textwidth}
         \centering
         \includegraphics[width=\textwidth]{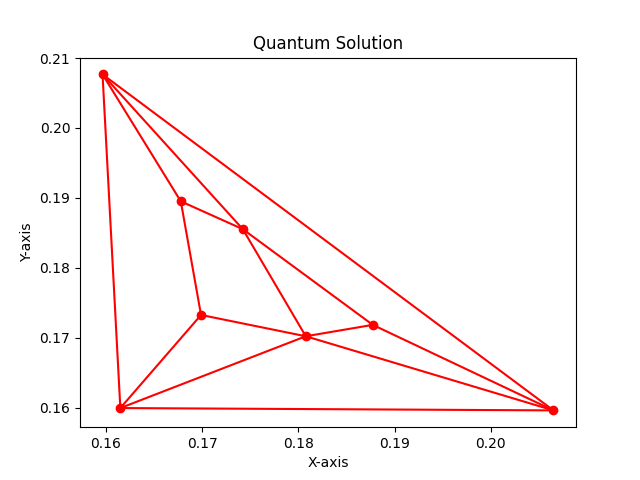}
     \end{subfigure}
     \hfill

     \begin{subfigure}[b]{0.49\textwidth}
         \centering
         \includegraphics[width=\textwidth]{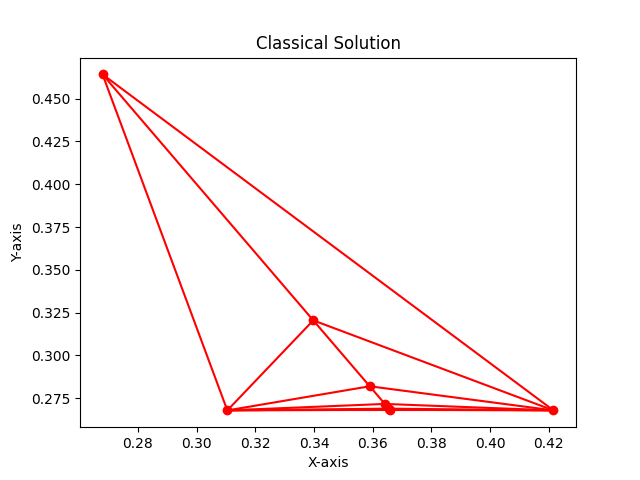}
     \end{subfigure}
     \hfill
     \begin{subfigure}[b]{0.49\textwidth}
         \centering
         \includegraphics[width=\textwidth]{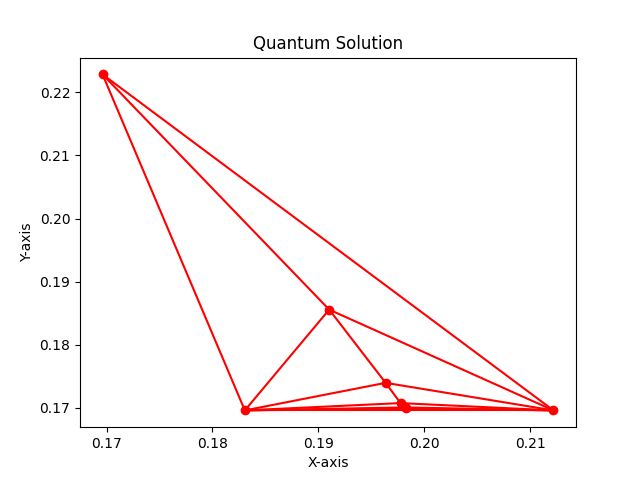}
     \end{subfigure}
     \hfill
     
     \begin{subfigure}[b]{0.49\textwidth}
         \centering
         \includegraphics[width=\textwidth]{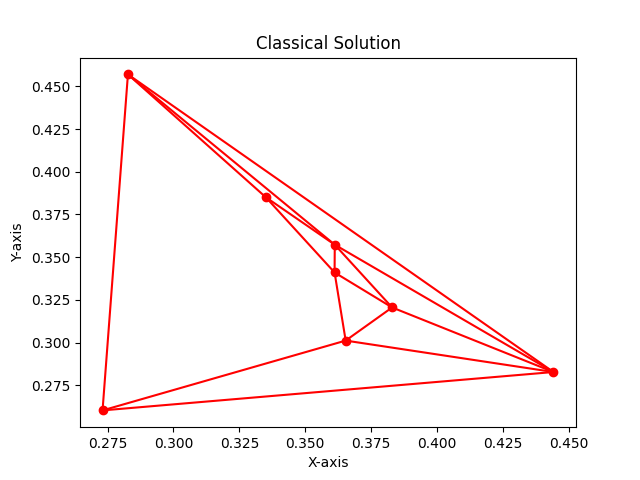}
     \end{subfigure}
     \hfill
     \begin{subfigure}[b]{0.49\textwidth}
         \centering
         \includegraphics[width=\textwidth]{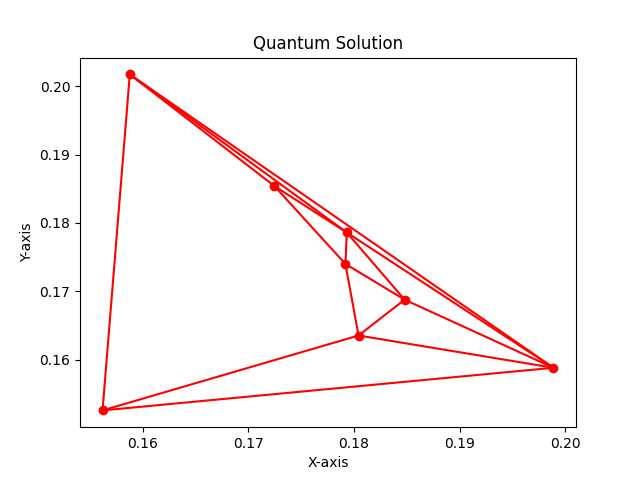}
     \end{subfigure}
     \hfill
    \caption{Graph drawing results using both the classical and quantum solver. The first is drawing a random planar graph with 4 nodes, and the last three are drawing random planar graphs with 8 nodes. }
    \label{fig:drawn-graphs-1}
\end{figure}
\begin{figure}
\centering
     \begin{subfigure}[b]{0.49\textwidth}
         \centering
         \includegraphics[width=\textwidth]{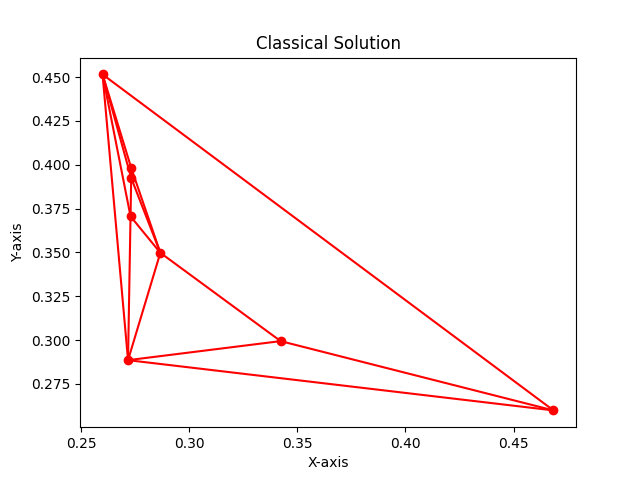}
     \end{subfigure}
     \hfill
     \begin{subfigure}[b]{0.49\textwidth}
         \centering
         \includegraphics[width=\textwidth]{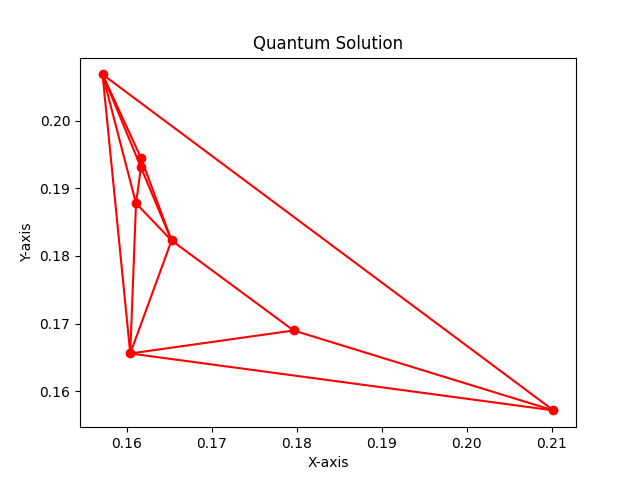}
     \end{subfigure}
     \hfill

          \begin{subfigure}[b]{0.49\textwidth}
         \centering
         \includegraphics[width=\textwidth]{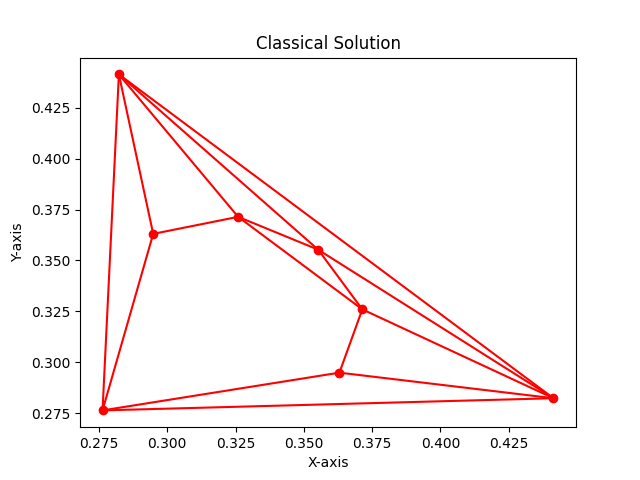}
     \end{subfigure}
     \hfill
     \begin{subfigure}[b]{0.49\textwidth}
         \centering
         \includegraphics[width=\textwidth]{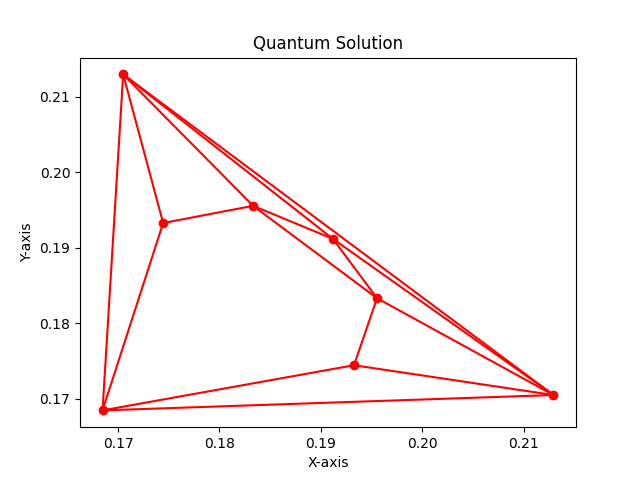}
     \end{subfigure}
     \hfill

     \begin{subfigure}[b]{0.49\textwidth}
         \centering
         \includegraphics[width=\textwidth]{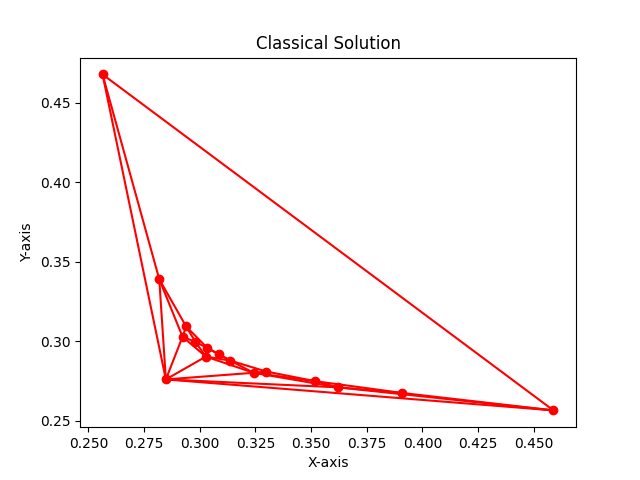}
     \end{subfigure}
     \hfill
     \begin{subfigure}[b]{0.49\textwidth}
         \centering
         \includegraphics[width=\textwidth]{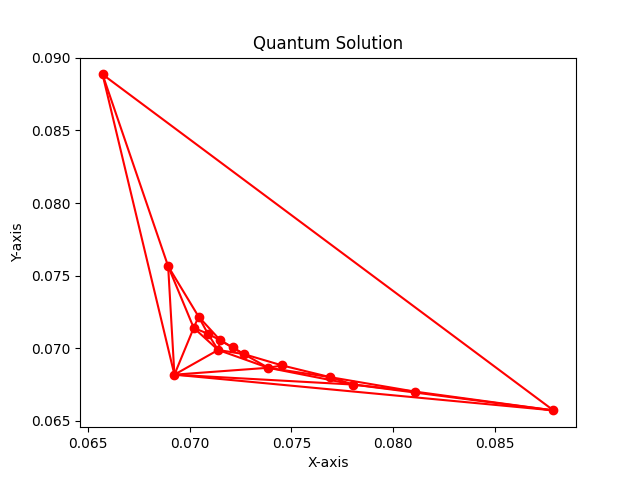}
     \end{subfigure}
     \hfill

     \begin{subfigure}[b]{0.49\textwidth}
         \centering
         \includegraphics[width=\textwidth]{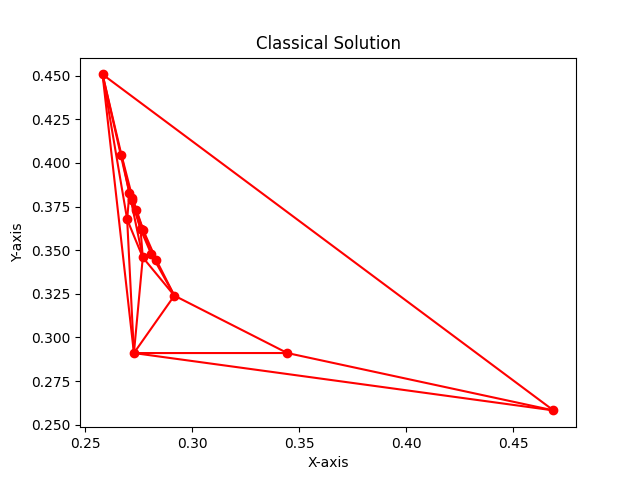}
     \end{subfigure}
     \hfill
     \begin{subfigure}[b]{0.49\textwidth}
         \centering
         \includegraphics[width=\textwidth]{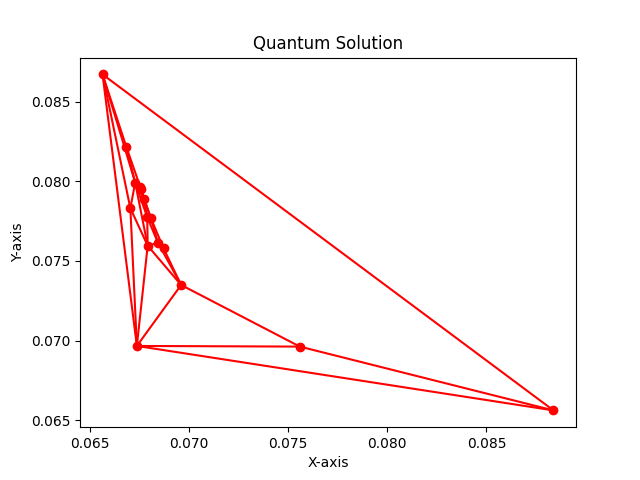}
     \end{subfigure}
     \hfill

    \caption{Graph drawing results using both the classical and quantum solver. The first two are drawing random planar graphs with 8 nodes, and the last two are drawing random planar graphs with 16 nodes. }
    \label{fig:drawn-graphs-2}
\end{figure}

To demonstrate the proof of concept for our Tutte embedding procedure
using quantum circuits, we used
a public 
implementation~\footnote{\url{https://github.com/anedumla/quantum_linear_solvers/tree/main}}
of Harrow's algorithm, which is endorsed by and works
with QISKit~\cite{cross2018ibm}, 
the leading software package for writing quantum circuits. 
We note that the off-the-shelf implementation provided
here does not perform the optimal matrix-to-circuit construction
as outlined in section 3. However, we adopt this version for the
purpose of demonstration. Furthermore, we use the circuit simulator
and thus only demonstrate the results for small-size graphs. The
time required for running the quantum simulator grows exponentially
with the number of qubits, and even for a graph with 16 nodes, this
required using matrices with size $2^{13}$ by $2^{13}$. This makes
simulating the quantum solver for larger-size graphs very difficult.
Some of the resulting graphs are shown in Figures \ref{fig:drawn-graphs-1}
and \ref{fig:drawn-graphs-2}.

The graphs drawn were generated randomly by placing $N$ random points on the plane, then defining their edges via a Delaunay triangulation. This guarantees their planarity as well as strong connectivity while maintaining low average degree per vertex. The dummy outer face is removed from the final drawing, leading to graphs whose outer faces are not perfectly placed triangles as their coordinates were not fixed.

%% file: appendix.tex
\section{A Brief Primer on Quantum Computing}

\begin{table}[]
\centering
\begin{tabular}{|l|c|}
\hline
Gate                      & Matrix Representation \\ \hline
$I$ (Identity)            & $\begin{bmatrix} 1 & 0 \\ 0 & 1 \end{bmatrix}$                              \\[0.3cm]
$X$ (Bitflip)             & $\begin{bmatrix} 0 & 1 \\ 1 & 0 \end{bmatrix}$                                \\[0.3cm]
$Z$                       & $\begin{bmatrix} 1 & 0 \\ 0 & -1 \end{bmatrix}$                             \\[0.3cm]
$H$                       & $\frac{1}{\sqrt{2}} \begin{bmatrix} 1 & 1 \\ 1 & -1 \end{bmatrix}$             \\[0.3cm]
$P(\theta)$               & $\begin{bmatrix} 1 & 0 \\ 0 & e^{i\theta} \end{bmatrix}$\\[0.3cm]
$CX$ (Controlled Bitflip) & $\begin{bmatrix} I & \\ & X \end{bmatrix}$                                    \\[0.3cm]
Toffoli                   & $\begin{bmatrix} I & & & \\ & I & & \\ & & I & \\ & & & X \end{bmatrix}$     \\ 
\hline
\end{tabular}
\caption{The matrix representation of quantum gates which will appear in this paper. The final two matrices are block encoded using the I and X single qubit gates. The controlled bitflip applies an X gate to the second qubit if the first qubit is set to 1. The Toffoli can be thought of as a doubly controlled bitflip, applying an X gate to a third qubit iff the first two qubits are set to 1. }
\label{tab:quantum-gates}
\end{table}

In this appendix, we go over some notation and properties of quantum computing 
that are important in describing quantum algorithms. For a full introduction, we refer the reader to Nielsen and Chuang~\cite{nielsen2010}. 

In a classical computer, an $n$-bit register can only store a single $n$-bit string at a time, but a quantum computer with an $n$-qubit register can store a  \textbf{superposition} of $n$-bit strings. This is represented as a unit column vector referred to as a \emph{state vector}, $\ket{\psi} = \sum_{i=0}^{n-1} \alpha_i \ket{i}$ (read "ket" psi), and the $\alpha_i$'s are complex numbers such that their norm squared is equal to the probability of measuring the bit string $i$. Thus, the quantum states must always be a unit vector whose  $L_2$-norm $\sum_i |\alpha_i|^2$ equals $1$.
Quantum gates are represented as unitary matrices that act on state vectors, effectively altering the superposition at each application. Unitarity guarantees that the state vector remains a unit vector after the transformation. 

An important operation in quantum computing is measurement. Although it is possible to store a superposition of states 
and operate on them concurrently, to read out any information about the state vector a measurement must be performed. The probably of measuring the bit string $i$ is equal to $|\alpha_i|^2$, and once measured, the state vector \textbf{collapses} to the bit string $i$, losing information about the state it collapsed from.\footnote{Although it is possible to perform different measurements, for simplicity we focus on measurement in the ``standard basis'' which measures the qubits as classical bit strings. }
Since the output of a measurement is drawn from a probability distribution determined by the squares of the
amplitudes of the output state, quantum computation is inherently probabilistic in nature. The goal
is to output a quantum state 
such that the ``solution'' bit string is measured with high probability. The probability of success can be boosted by repetition as with classical randomized algorithms. 

\section{Example Source Code}

We provide in Figure~\ref{fig:code} example source code for our
quantum simulations.

\begin{figure}
{\small
\begin{verbatim}
import numpy as np
import scipy
from linear_solvers import HHL
import random
from qiskit.quantum_info import Statevector

# Number of vertices
n = 8

# Generate a grounded Laplacian for input graph
L = Input graph Laplacian 
A = L
A[0][0] += 1
A[1][1] += 1
A[2][2] += 1

b_x = np.array([0, 0, 1, 0, 0, 0, 0, 0])
b_y = np.array([0, 1, 0, 0, 0, 0, 0, 0])

def get_solution_vector(solution):
    """Extracts and normalizes simulated state vector
    from LinearSolverResult."""

    # Read out the relevant entries in the state vector
    half_dim = Statevector(solution.state).dim // 2
    solution_vector = Statevector(solution.state).data[half_dim:half_dim+n].real
    print(Statevector(solution.state).dim)
    
    return solution_vector

# Instantiate HHL object and solve for x and y coordinates.
hhl = HHL(epsilon=1e-4)
naive_hhl_solution_x = hhl.solve(A, b_x)
sol_x = get_solution_vector(naive_hhl_solution_x) * hhl.scaling
hhl = HHL(epsilon=1e-4)
naive_hhl_solution_y = hhl.solve(A, b_y)
sol_y = get_solution_vector(naive_hhl_solution_y) * hhl.scaling
\end{verbatim}
}
\caption{Example source code for our quantum simulation.}
\label{fig:code}
\end{figure}

%% file: paper.bbl
\begin{thebibliography}{10}
\providecommand{\url}[1]{\texttt{#1}}
\providecommand{\urlprefix}{URL }

\bibitem{bassoli2021quantum}
Bassoli, R., Boche, H., Deppe, C., Ferrara, R., Fitzek, F.H., Janssen, G.,
  Saeedinaeeni, S.: Quantum Communication Networks, vol.~23. Springer (2021)

\bibitem{Cabello_2012}
Cabello, A., Danielsen, L.E., L{\'o}pez-Tarrida, A.J., Portillo, J.R.: Quantum
  social networks. Journal of Physics A: Mathematical and Theoretical  45(28),
  285101 (jun 2012), \url{https://dx.doi.org/10.1088/1751-8113/45/28/285101}

\bibitem{childs2017}
Childs, A.M., Kothari, R., Somma, R.D.: Quantum {{Algorithm}} for {{Systems}}
  of {{Linear Equations}} with {{Exponentially Improved Dependence}} on
  {{Precision}}. SIAM Journal on Computing  46(6),  1920--1950 (Jan 2017)

\bibitem{chung1997spectral}
Chung, F.R.: Spectral Graph Theory, vol.~92. American Mathematical Soc. (1997)

\bibitem{cross2018ibm}
Cross, A.: The {IBM Q} experience and {QISKit} open-source quantum computing
  software. In: APS March Meeting Abstracts. vol. 2018, pp. L58--003 (2018)

\bibitem{dervovic2018}
Dervovic, D., Herbster, M., Mountney, P., Severini, S., Usher, N., Wossnig, L.:
  Quantum linear systems algorithms: A primer (Feb 2018)

\bibitem{battista1999graph}
{Di Battista}, G., Eades, P., Tamassia, R., Tollis, I.G.: Graph Drawing:
  {Algorithms} for the Visualization of Graphs. Prentice Hall (1999)

\bibitem{istvan1948straight}
F{\'a}ry, I.: On straight-line representation of planar graphs. Acta
  Scientiarum Mathematicarum  11(2),  229--233 (1948)

\bibitem{Floater}
Floater, M.S.: Parametric tilings and scattered data approximation.
  International Journal of Shape Modeling  04(03n04),  165--182 (1998)

\bibitem{harrow2009}
Harrow, A.W., Hassidim, A., Lloyd, S.: Quantum algorithm for linear systems of
  equations. Phys. Rev. Lett.  103,  150502 (Oct 2009),
  \url{https://link.aps.org/doi/10.1103/PhysRevLett.103.150502}

\bibitem{hopcroft1992paradigm}
Hopcroft, J.E., Kahn, P.J.: A paradigm for robust geometric algorithms.
  Algorithmica  7(1-6),  339--380 (1992)

\bibitem{kelner2013simple}
Kelner, J.A., Orecchia, L., Sidford, A., Zhu, Z.A.: A simple, combinatorial
  algorithm for solving {SDD} systems in nearly-linear time. In: 45th ACM
  Symposium on Theory of Computing (STOC). pp. 911--920 (2013)

\bibitem{kitaev95}
Kitaev, A.Y.: Quantum measurements and the abelian stabilizer problem (1995)

\bibitem{kobourov2013}
Kobourov, S.G.: Force-directed drawing algorithms. In: Tamassia, R. (ed.)
  Handbook of Graph Drawing and Visualization, chap.~12, pp. 383--408. CRC
  Press (2013)

\bibitem{kumar2021quantum}
Kumar, R., Kumari, S., Bala, M.: Quantum mechanical model of information
  sharing in social networks. Social Network Analysis and Mining  11(1), ~42
  (2021)

\bibitem{matta2010quantum}
Matta, C.F.: Quantum Biochemistry. John Wiley \& Sons (2010)

\bibitem{nielsen2010}
Nielsen, M.A., Chuang, I.: Quantum Computation and Quantum Information.
  Cambridge University Press, 10th edn. (2010)

\bibitem{pirani2015}
Pirani, M., Sundaram, S.: On the {{Smallest Eigenvalue}} of {{Grounded
  Laplacian Matrices}}. IEEE Transactions on Automatic Control pp. 1--1 (2015)

\bibitem{spielman2004nearly}
Spielman, D.A., Teng, S.H.: Nearly-linear time algorithms for graph
  partitioning, graph sparsification, and solving linear systems. In: 36th ACM
  Symposium on Theory of Computing (STOC). pp. 81--90 (2004)

\bibitem{stein1951convex}
Stein, S.K.: Convex maps. Proceedings of the American Mathematical Society
  2(3),  464--466 (1951)

\bibitem{trotter1959product}
Trotter, H.F.: On the product of semi-groups of operators. Proceedings of the
  American Mathematical Society  10(4),  545--551 (1959)

\bibitem{tutte1963draw}
Tutte, W.T.: How to draw a graph. Proceedings of the London Mathematical
  Society  3(1),  743--767 (1963)

\bibitem{wagner1936bemerkungen}
Wagner, K.: Bemerkungen zum vierfarbenproblem. Jahresbericht der Deutschen
  Mathematiker-Vereinigung  46,  26--32 (1936)

\end{thebibliography}
